\documentclass[reprint, amsmath, aps]{revtex4-1}

\usepackage{hyperref}
\usepackage{graphicx}
\usepackage{xcolor}

\hypersetup{colorlinks, linkcolor=blue, citecolor=blue, urlcolor=blue}

\newcommand{\colscale}{0.9\columnwidth}
\newcommand{\textscale}{0.9\textwidth}

\newcommand{\PT}{\mathcal{PT}}
\newcommand{\textPT}{\texorpdfstring{$\PT$}{PT}}
\newcommand{\bra}[1]{\langle{#1}|}
\newcommand{\ket}[1]{|{#1}\rangle}

\begin{document}

\title{Real Edge Modes in a Floquet-modulated \textPT-symmetric SSH Model}

\author{Andrew K. Harter}
\author{Naomichi Hatano}

\affiliation{Institute of Industrial Science, The University of Tokyo \\
5-1-5 Kashiwanoha, Kashiwa \\
Chiba 277-8574, Japan}

\date{\today}

\begin{abstract}
    Non-Hermitian Hamiltonians provide a simple picture for analyzing systems with natural or induced gain and loss; however, in general, such Hamiltonians feature complex energies and a corresponding non-orthonormal eigenbasis. Provided that the Hamiltonian has $\PT$ symmetry, it is possible to find a regime in which the eigenspectrum is completely real. In the case of static $PT$-symmetric extensions of the simple Su-Schrieffer-Heeger model, it has been shown that the energies associated with any edge states are guaranteed to be complex. Moving to a time-dependent system means that treatment of the Hamiltonian must be done at the effective time-scale of the modulation itself, allowing for more intricate phases to occur than in the static case. It has been demonstrated that with particular classes of periodic driving, achieving a real topological phase at high driving frequency is possible. In the present paper, we show the details of this process by using a simple two-step periodic modulation. We obtain a rigorous expression for the effective Floquet Hamiltonian and compare its symmetries to those of the original Hamiltonians which comprise the modulation steps. The $\PT$ phase of the effective Hamiltonian is dependent on the modulation frequency as well as the gain/loss strength. Furthermore, the topologically nontrivial regime of the $\PT$-unbroken phase admits highly-localized edge states with real eigenvalues in both the high frequency case and below it, albeit within a smaller extent of the parameter space.
\end{abstract}

\maketitle

\section{Introduction} \label{sec:introduction}

Parity-time ($\PT$) symmetry \cite{Bender1998} was first introduced as a fundamentally non-Hermitian extension of quantum mechanics \cite{Bender2002} and is now often used to model open systems \cite{El-Ganainy2007, Guo2009, Ruter2010} with symmetrically balanced loss and gain. The primary feature of $\PT$-symmetric systems is that up to a critical strength of gain and loss, the non-Hermitian system Hamiltonian has all real eigenvalues, and it shares its eigenstates with the $\PT$ operator. Above this threshold, pairs of complex conjugate eigenvalues form, breaking the $\PT$ symmetry.

Static systems with $\PT$ symmetry have been studied for the past several decades \cite{Feng2017, El-Ganainy2018, Ozdemir2019}, and have proven to be experimentally achievable in a wide variety of settings including waveguides \cite{Guo2009, Ruter2010, Zeuner2015, Kremer2019}, optical resonators \cite{Hodaei2014, Hodaei2016}, electrical circuits \cite{Schindler2011}, mechanical systems \cite{Bender2013}, and atomic systems \cite{Zhang2016}. While these static systems provide ample grounds to explore many of the basic properties of $\PT$-symmetric systems, they have several drawbacks. Chiefly, for extended systems, gain is difficult to implement in a way which exactly balances the loss, which often has its origins in the natural coupling of the system to its environment. For this reason, it is often convenient to transform a $\PT$-symmetric system to a loss-only one, which can be more easily implemented. Experimentally, large arrays of optical waveguides have proven to be useful in the study of these ``lossy'' $\PT$-symmetric systems \cite{Zeuner2015, Kremer2019}.

Building on this theory, a lattice model was proposed by Rudner and Levitov \cite{Rudner2009} to study topology in a $\PT$-symmetric system. This model is a natural extension of the Hermitian Su-Schrieffer-Heeger (SSH) model \cite{Su1979}, a simple one-dimensional dimer lattice which exhibits a well-known topological transition and topologically protected mid-gap edge states \cite{Asboth2016}. The model of Rudner and Levitov also exhibits many of these qualities; however, the associated energies of the localized edge states are pure imaginary, so that they are dynamically unstable \cite{Hu2011}. Despite this, the topological transition in this model is well-defined \cite{Liang2013, Harter2018, Lieu2018} and has been experimentally observed \cite{Zeuner2015}.

In the present paper, we study a time-periodic (Floquet) system \cite{Shirley1965} which is fashioned after that of Rudner and Levitov \cite{Rudner2009}, but with a modulated gain and loss rate \cite{Joglekar2014, Lee2015, Li2019, Leon-Montiel2018}. A similar system has previously been shown to exhibit an entirely real spectrum in the high-driving-frequency regime \cite{Yuce2015}. According to the Floquet theory, the long-time dynamics of such a time-dependent system cannot be determined by examining a single static system; rather, we may introduce an effective static system, which is determined by the evolution over one period of the modulation \cite{Shirley1965, Kitagawa2010}. The introduction of the periodic time dependence allows states which are dynamically unstable in the static case to be stabilized by the periodic modulation while introducing a novel topology unique to Floquet systems \cite{Rechtsman2013, Rudner2013}.

We begin, in Sec.~\ref{sec:static}, with a quick overview of the topologically relevant, Hermitian SSH model in Sec.~\ref{sec:ssh}, after which we discuss, in Sec.~\ref{sec:pt-ssh}, the static model proposed by Rudner and Levitov \cite{Rudner2009} which is its extension. We also provide a quick discussion of the analysis of this model in momentum space in Sec.~\ref{sec:momentum-space}. In Sec.~\ref{sec:floquet-pt}, we introduce the Floquet driving of the $\PT$-SSH model. We begin with a discussion of the Floquet theory in Sec.~\ref{sec:floquet}, followed by an analytical treatment in Sec.~\ref{sec:analytical-floquet}, and an analysis of the edge states in Sec.~\ref{sec:edge-states}. Finally, in Sec.~\ref{sec:conclusion}, we conclude with a brief discussion of the merits of our model.

\section{The Static \textPT-SSH Model} \label{sec:static}

We first briefly review the relevant results of the static $\PT$-SSH model. To understand this non-Hermitian model, we begin with the Hermitian SSH model \cite{Su1979} in Sec.~\ref{sec:ssh}, which is a simple dimer lattice with alternating couplings. Then, in Sec.~\ref{sec:pt-ssh}, we add to this a $\PT$-symmetric extension \cite{Rudner2009, Zeuner2015} which attempts to respect the lattice symmetry.

\subsection{The SSH Model} \label{sec:ssh}

The starting point for our study is the Su-Schrieffer-Hegger (SSH) model \cite{Su1979}, a one-dimensional dimer chain, the bulk of which is described by the Hermitian tight-binding Hamiltonian
\begin{equation}\label{eqn:bulk-ssh}
    H_{SSH} = \sum_m\left(v\ket{m,A}\bra{m,B} + w\ket{m,B}\bra{m+1,A}\right) + \mathrm{h.c.}
\end{equation}
with intra-dimer coupling $v$ and inter-dimer coupling $w$, both real-valued quantities. Each dimer cell is labeled $m$ with sub-lattice labels $A$ and $B$. We further impose open boundary conditions with $m$ ranging from $1$ to $M$ and the summation of Eqn.~(\ref{eqn:bulk-ssh}) cut off appropriately so that there are exactly $M$ complete dimers.

\begin{figure}
    \centering
    \includegraphics[width=\colscale]{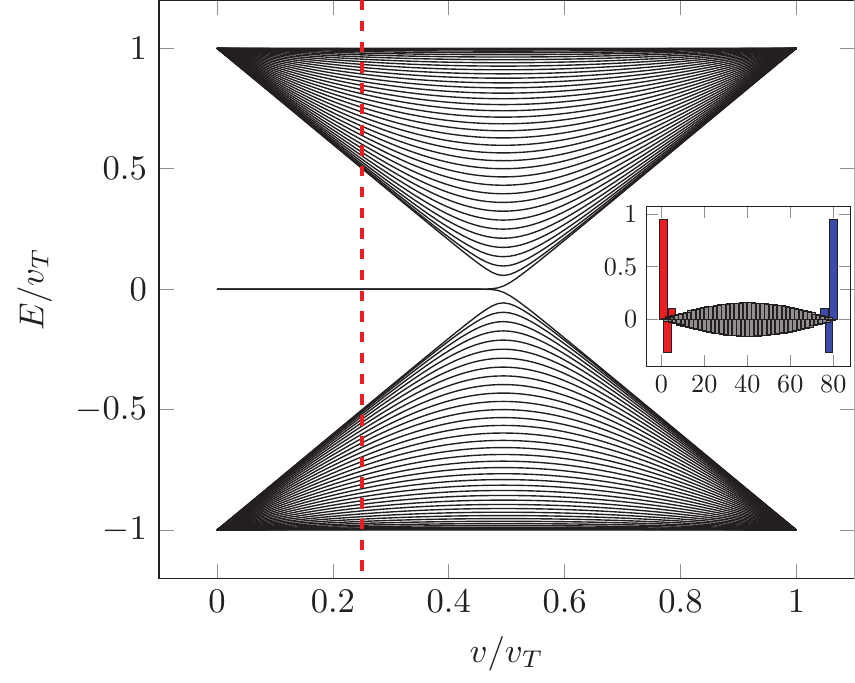}
    \caption{We show, for reference, the energy bands of the static SSH model plotted against the changing lattice parameter $v/v_T$. In the smaller inset, we show the wavefunction amplitudes of the two edge states with the eigenvalues $E = 0$ and a bulk state with $E = -J$ (plotted against each spatial site on the horizontal axis) for the parameterization corresponding to $v/v_t = 0.25$ (which is indicated by the vertical dashed red line in the larger plot). The edge states are highly localized to the left and right of the graph (red and blue), while the amplitude profile for a bulk state is highly delocalized (dark-gray).}
    \label{fig:ssh-bands}
\end{figure}

We explore parameterizations of this model in terms of its intra-dimer coupling $v$ by defining the scale of relevant coupling frequencies $v_T \equiv v + w$. Specifically, along the parameter range $v/v_T \in [0, 1]$, the couplings range between two fully dimerized limits $(v, w) = (0, v_T)$ and $(v, w) = (v_T, 0)$.

In Fig.~\ref{fig:ssh-bands}, we see that the bulk energy bands are gapped by $\Delta E = 2|v - w|$, so that at exactly $v = w$, namely the uniform lattice, the gap closes. Due to the topological nature of the lattice, when $v < w$, or equivalently $v/v_T < 0.5$, the boundary condition of the lattice introduces a pair of protected edge states (see the inset of Fig.~\ref{fig:ssh-bands}), and the lattice exhibits a topologically nontrivial phase, as can be seen by the presence of the mid-gap zero-energy eigenvalue. However, when $v > w$, or equivalently, $v/v_T > 0.5$, these states are absorbed into the bulk.

At the dimerization limits $v/v_T \rightarrow 0$ and $v/v_T \rightarrow 1$, we can easily inspect the topological nature of the real-space Hamiltonian. In the case of $v = 0$, both edges of the system are completely decoupled from the bulk of the system, resulting in two zero-energy eigenvalues given by the states $\ket{1,A}$ and $\ket{M,B}$, which are exactly localized on either edge of the system. In general, moving away from this limit, the edge states remain localized, as is demonstrated in the inset of Fig.~\ref{fig:ssh-bands}, until the gap closes at exactly $v/v_T = 0.5$. In the other limit, when $v/v_T = 1$ ($w = 0$), the unit cells are completely decoupled from each other. Its eigenstates are given by the direct products of independent dimers, and hence, no zero-energy edge states exist. Away from this limit, the dimers couple with each other, but the edge states remain nonexistent until the gap closes at $v/v_T = 0.5$.

\subsection{The \textPT-Extended SSH Model} \label{sec:pt-ssh}

\begin{figure}
    \centering
    \includegraphics[width=\colscale]{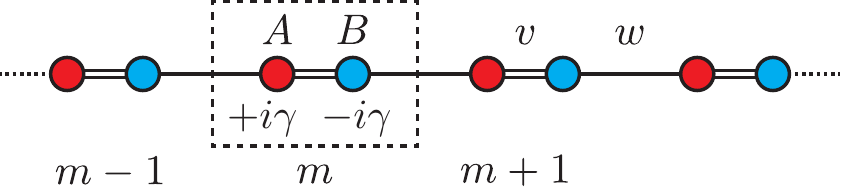}
    \caption{Depiction of the $\PT$-SSH lattice with intra-dimer coupling $v$ and inter-dimer coupling $w$. The gain site ($+i\gamma$) is indicated by the red color, and the loss site ($-i\gamma$) is indicated by the blue color. Each pair of sites A and B (red and blue) comprises a full unit cell (labeled by $m$). Inside the cell, the sites are coupled by $v$, and the cells themselves are coupled by the alternate parameter $w$. We define the coupling scale by $v_T = v + w$.}
    \label{fig:pt-ssh-lattice}
\end{figure}

A non-Hermitian extension of this model was proposed by Rudner and Levitov \cite{Rudner2009}. In this model, an additional $\PT$-symmetric component which is local to each dimer is added to the original SSH model,
\begin{equation}\label{eqn:pt-ssh}
    H_{\PT} = H_{SSH} + i\gamma\sum_m \left( \ket{m,A}\bra{m,A} - \ket{m,B}\bra{m,B} \right) \,,
\end{equation}
as depicted in Fig.~\ref{fig:pt-ssh-lattice}, where $\gamma$ is a real parameter that controls the gain and loss rate in the system.

\begin{figure*}
    \centering
    \includegraphics[width=\textscale]{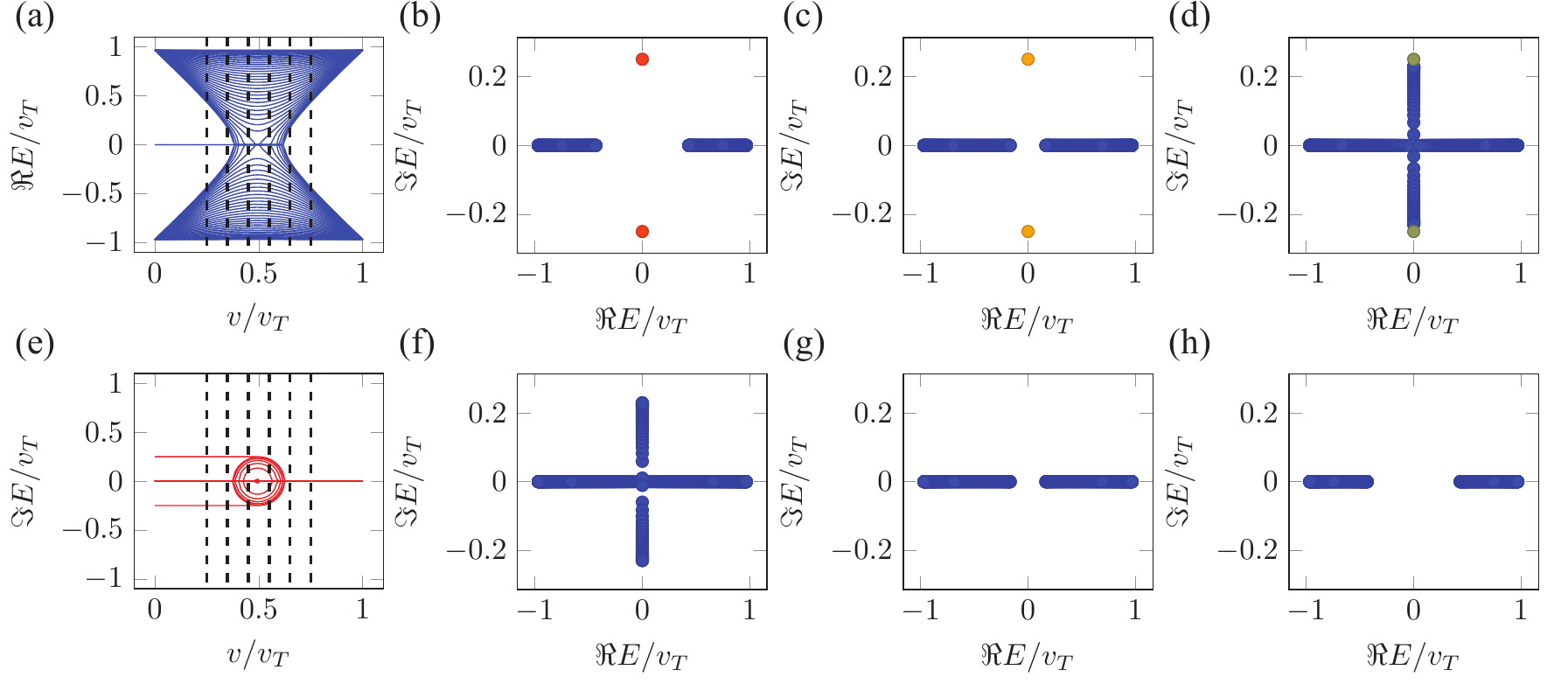}
    \caption{Depiction of the complex energy spectrum of the $\PT$-extended SSH model with fixed gain/loss rate $\gamma/v_T = 0.25$. In (a) and (e), we show the real and imaginary parts, respectively, of the energy spectrum as a function of the configuration parameter $v/v_T$. The vertical dashed lines correspond to the six values of $v/v_T$ highlighted by (b)-(d) and (f)-(h). In panels (b)-(d), we show the complex energy spectrum of the Hamiltonian for configurations having $v/v_T < 0.5$; each complex energy corresponds to a complex coordinate which is colored according to its IPR (from blue at zero to red at one). Specifically $v/v_T = 0.25$ in (b), $v/v_T = 0.35$ in (c), and $v/v_T = 0.45$ in (d). In panels (f)-(h), we show the complex energy spectrum for configurations having $v/v_T > 0.5$, with $v/v_T = 0.55$ in (f), $v/v_T = 0.65$ in (g), and $v/v_T = 0.75$ in (h).}
    \label{fig:pt-ssh-bands}
\end{figure*}

In this case, the Hamiltonian is no longer Hermitian, and the energy eigenvalues may take on imaginary components. In Fig.~\ref{fig:pt-ssh-bands}, we show the energies in the complex plane as a function of $v/v_T$ \cite{Lieu2018}. In Fig.~\ref{fig:pt-ssh-bands} (b)-(d), we observe the presence of non-bulk states (red), although they are pure imaginary. Between (c) and (d), the bulk states pick up imaginary components and begin to extend towards the isolated states on the imaginary axis. In Fig.~\ref{fig:pt-ssh-bands} (f)-(h), we observe the recession of the imaginary bulk states from (f) to (g); in (g), a gap begins to open between the purely real bulk bands, but the edge states are removed.

In order to see whether the states are localized or extended, we can also calculate the inverse participation ratio (IPR) for each state; the IPR, defined by
\begin{equation}
    \mathrm{IPR}(\psi) = \sum_m |\psi_m|^4 \,,
\end{equation}
is a measure of the extent to which a given state is localized. For the states indicated by blue dots in Fig.~\ref{fig:pt-ssh-bands} (f)-(h), the IPR is small $\leq 0.02$, whereas for the states with dark gray dots in (d), the IPR is $\approx 0.20$, and for the edge states indicated by red/yellow states in (b) and (c), the IPR is large $\approx 0.55$ (yellow) in (c) and $\approx 0.80$ (red) in (b). It is immediately clear that the localized modes are the ones which lie in the imaginary plane and they are complex conjugates.

Moving away from these limits, to the full domain of $v$ and $w$, this model was found (under the appropriate imaginary guage transformation) to admit a topological invariant which is quantized to $1$ for $v/v_T < 0.5$, and $0$ for $v/v_T > 0.5$ \cite{Rudner2009, Liang2013}. This invariant consists of a global integration over space and time, and is not associated to the upper or lower bands as is the case in the Hermitian SSH model \cite{Rudner2009, Lieu2018}. While there has been much discussion \cite{Rudner2009, Liang2013, Lieu2018, Kowabata2019} as to the nature of the topology of this and other non-Hermitian models, we focus our effort on the presence (or absence) of dynamically stable edge states.

In light of this, the $\PT$-extended SSH model does not have a topological phase with dynamically stable edge states \cite{Hu2011}; when $v/v_T < 0.5$, the edge states are inherently robust against perturbations in $v$, and separately, when $\gamma < |v - w|$, where the energy spectrum is entirely real (possessing unbroken $\PT$ symmetry), the system is dynamically stable; however, these two features do not coexist in this extended model. In the case where there are localized edges states, we can see that the system always has a pair of imaginary eigenvalues which correspond to the topological states; and thus the topological states are inherently dynamically unstable.

\subsection{Momentum Space Analysis} \label{sec:momentum-space}

Considering just the bulk of the system, let us neglect the boundaries for the moment. For a given intra-dimer coupling $v$ and gain/loss strength $\gamma$, the bulk Hamiltonian is block-diagonal in momentum space, with each block corresponding to a value $k \in [-\pi, \pi)$ having the form
\begin{equation} \label{eqn:hptk}
    H_\PT(k) = (v + w\cos k)\sigma_x + \left(w\sin k\right)\sigma_y + i\gamma\sigma_z \,,
\end{equation}
where $\sigma_i$ are the Pauli matrices. Therefore, each block in momentum space effectively represents a two-site system with a complex, $k$-dependent coupling. Furthermore, in a rotated frame which changes with $k$, we may write
\begin{equation} \label{eqn:hptkrot}
    H'_\PT(k) = r(k)\sigma_x + i\gamma\sigma_z \,,
\end{equation}
where
\begin{equation}\label{eqn:rk}
    r(k) = \sqrt{v^2 + w^2 + 2vw\cos k} \,
\end{equation}
is real and takes values between $|v-w|$ and $v + w$ over the range of $k$. The rotation of the frame is given by
\begin{equation} \label{eqn:krotation}
H_\PT(k) = e^{-i\sigma_z\phi(k)/2}H'_\PT(k) e^{i\sigma_z\phi(k)/2} \,,
\end{equation}
where we have defined the angle
\begin{equation}
    \phi(k) = \tan^{-1} \frac{w\sin k}{v + w\cos k} \,.
\end{equation}
The eigenvalues of the matrix of Eq.~(\ref{eqn:hptkrot}) are $E(k) = \sqrt{r^2(k) - \gamma^2}$, which indicates that when $\gamma \leq |v - w|$, $E(k)$ is real over the entire range of $k$.

Thus, we may determine the time development under the real-space matrix $H_\PT$ in terms of the time development of $H'_\PT(k)$ and we focus our attention to the $k$-space evolution. The time-evolution operator for the $k$th rotated block reduces to
\begin{align}
    \nonumber G_\PT(k,t) &= e^{-iH'_\PT(k)t} \\
    &= \cos[E(k)t]\,I - i\sin[E(k)t]\frac{H'_\PT(k)}{E(k)} \,,
\end{align}
where $I$ is the $2 \times 2$ identity matrix. It is important to note that the quantities $\cos[E(k)t]$ and $\sin[E(k)t] / E(k)$ are always real even when $\gamma > r(k)$ because $E(k)^2$ is real.

\section{The Floquet-Driven \textPT-SSH model} \label{sec:floquet-pt}

It has previously been discovered by Yuce \cite{Yuce2015} that periodic temporal (Floquet) modulation can provide a means to completely restore stability to the topologically nontrivial phase in the high-driving frequency regime. In this paper, we propose a concrete, exactly-solvable, two-step modulation which will allow us to gain insights into this process of stabilization and explore regions of stability below the high-driving frequency. Furthermore, utilizing the simplicity of the model, we seek to identify the underlying symmetry. This setup has also proved to be experimentally beneficial as it only requires control of the gain or loss in pulsed fashion \cite{Li2019}.

Consider the following two-step time sequence for $0 \leq t < T$, where $T$ is the period of modulation,
\begin{equation} \label{eqn:pt-pt-driving}
    H(t) = \begin{cases}
        H_\PT & \quad 0 \leq t < T/2 \,, \\
        \tilde{H}_\PT & \quad T/2 \leq t < T \,,
    \end{cases}
\end{equation}
with $H_\PT$ being the Hamiltonian (see Eq.~(\ref{eqn:pt-ssh})) of the static $\PT$-extended SSH lattice described in Sec.~\ref{sec:pt-ssh}, while $\tilde{H}_\PT$ is the Hamiltonian of $H_\PT$ with the sign of $\gamma$ inverted to $-\gamma$. This corresponds to a periodic modulation of the gain/loss strength which instantaneously interchanges the location of gain and loss, $\gamma$, at each step. In fact, we can say that $\tilde{H}_\PT$ is the time-reversed version of $H_\PT$, as in $\tilde{H}_\PT = \mathcal{T}H_\PT\mathcal{T} = H_\PT^*$.

\subsection{Floquet Analysis} \label{sec:floquet}

According to the Floquet Theorem, the time evolution operator of a periodically driven system can be written
\begin{equation}
    G(t) = P(t)e^{-iH_Ft} \,,
\end{equation}
where the matrix $P(t)$ is periodic with the same frequency as the driving, while $H_F$ is a constant matrix known as the effective Floquet Hamiltonian. Thus the long-term dynamics of the evolution are determined completely by $H_F$; further, at stroboscopic times $nT$, for any integer $n$, $P(nT)$ is the identity because $G(0) = P(0) = I$. Consequently, the system is dynamically unstable when at least one eigenvalue of $H_F(k)$ has an imaginary component, which results in exponential growth or decay in the long-term evolution of the system.

\begin{figure*}
    \centering
    \includegraphics[width=\textscale]{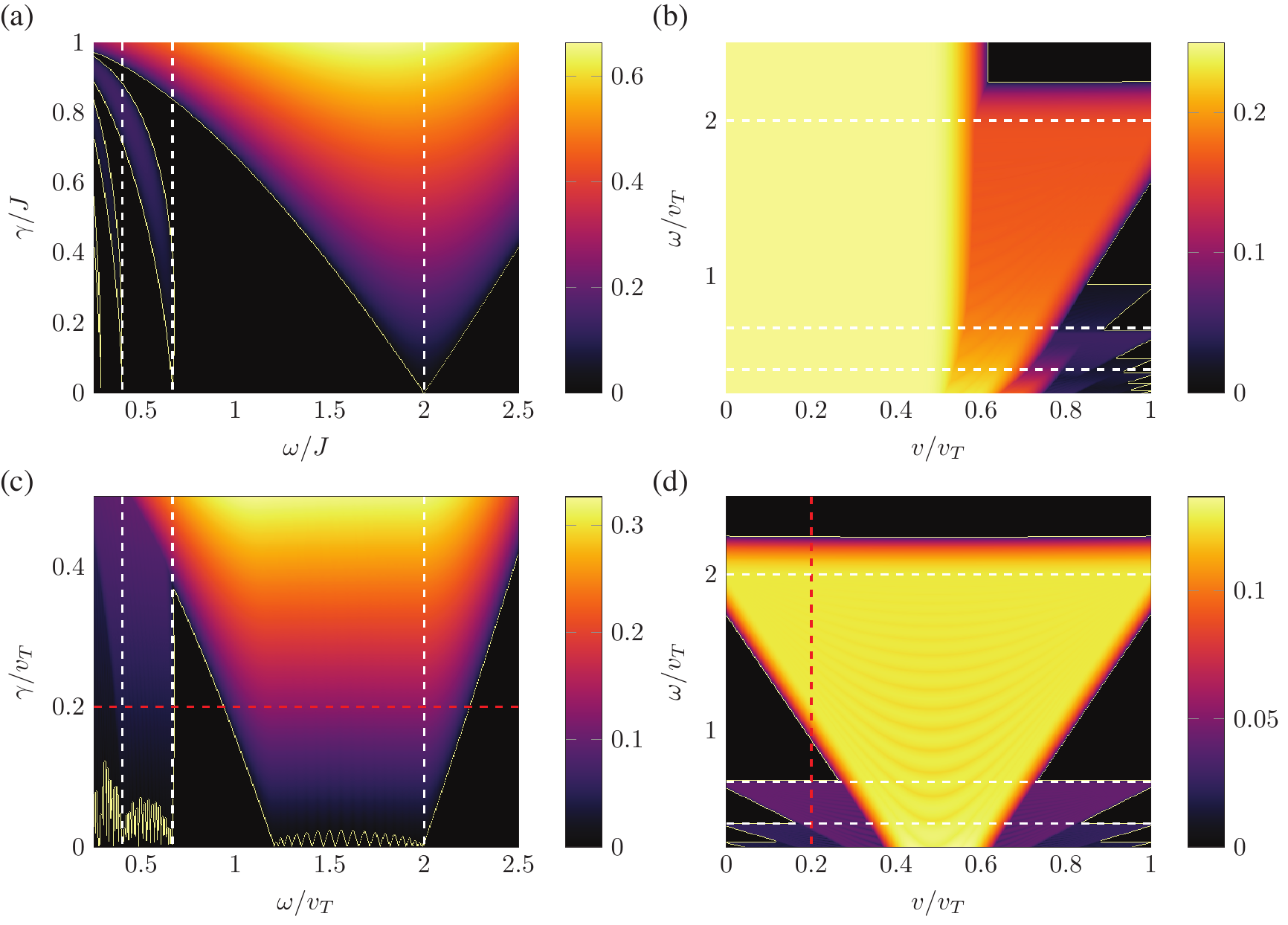}
    \caption{The $\PT$ phase diagrams for the Floquet $\PT$-symmetric systems discussed in Sec.~\ref{sec:floquet-pt}. In each figure, the magnitude of the maximum imaginary eigenvalue is plotted over a plane in parameter space, where the color indicates increasing magnitude (thus, the black regions indicate $\PT$-symmetric phases). As a visual aid, a solid line has been added which outlines the border between the $\PT$-symmetric and $\PT$-broken phases. (a) The phase diagram for the two-site system of Eq.~(\ref{eqn:hpt2}), over a range of gain/loss parameters $\gamma$ and driving frequencies $\omega$. (b) The phase diagram for the 40-site Floquet $\PT$-extended SSH driven between $\gamma/v_T = 0.5$ and $\gamma = 0$ over a range of driving frequencies and couplings $v/v_T$. (c) The phase diagram over the $\omega$-$\gamma$ plane as in (a), but for the 40-site Floquet $\PT$-extended SSH model driven by swapping the gain ($+i\gamma$) and loss ($-i\gamma$) sites and with fixed coupling $v/v_T = 0.2$. (d) The phase diagram for the same system of (c), but in the $v$-$\omega$ plane with gain/loss magnitudes fixed at $\gamma/v_T = 0.2$. The dashed red lines in (c) and (d) indicate a portion of the $\PT$-phase with equivalent parameter values. In each panel, the dashed white lines indicate locations of the first three resonances $\omega/v_T = 2/n$ ($\omega/J = 2/n$ in the two-site case) with $n = 1, 3, 5$. Note that there are additional resonances between these in the asymmetrically driven case of panel (b).}
    \label{fig:pt-phases}
\end{figure*}

For a two-site Hamiltonian
\begin{equation} \label{eqn:hpt2}
H_2 = J\sigma_x + i\gamma\sigma_z \,,
\end{equation}
where $J$ is a real coupling parameter, similar Floquet $\PT$-symetric models have previously been studied \cite{Joglekar2014, Lee2015} and found to exhibit a rich $\PT$-phase diagram featuring re-entrant $\PT$-symmetry breaking depending on the driving frequency $\omega \equiv 2\pi/T$. As is shown in Fig.~\ref{fig:pt-phases}(a), at specific driving frequencies, i.e. $\omega/J = 2, 2/3, 2/5, \ldots$, the $\PT$ symmetry of the system is broken for any nonzero value of $\gamma$ (i.e. the $\PT$-threshold is zero at these locations). The periodic modulation at these driving frequencies is in resonance with the natural coupling frequency of the two-site system, namely $2J$.

Utilizing the similarity between Eq.~(\ref{eqn:hptkrot}) and Eq.~(\ref{eqn:hpt2}), we can explore the $\PT$-phase diagram of the Floquet $\PT$-SSH model. In this case, as seen in Fig.~\ref{fig:pt-phases}(c), the resonance points occur for each value of $k$ at, i.e. $\omega/r(k) = 2, 2/3, 2/5, \ldots$, with $r(k)$ as defined in Eq.~(\ref{eqn:rk}), and there is an apparent broadening of the zero-$\PT$-threshold regime. In this case, the $\PT$-phase diagram can be thought of as the composition of the two-site case of Fig.~\ref{fig:pt-phases}(a) for multiple scaling values $r(k)$ determined by the allowed momenta $k$ replacing $J$ in Eq.~(\ref{eqn:hpt2}). The vertical dashed lines are placed at the resonance frequencies of the two-site case, demonstrating that they correspond to the two-site scaled system with the choice of $r(k=0) = v_T$. 

We can also analyze this model over a range of coupling parameters $v$ with a fixed value of gain/loss parameter $\gamma$, as is done in both Fig.~\ref{fig:pt-phases}(b), (d), where we show the $\PT$ phase in a sweep along the lattice parameter $v$ with driving frequencies $\omega$ increasing from bottom to top. For comparison to the previous two-site experimental studies \cite{Li2019, Leon-Montiel2018}, in Fig.~\ref{fig:pt-phases}(b), we modulate the system between a $\PT$-symmetric one ($\gamma > 0$) and a Hermitian one ($\gamma = 0$), instead of the modulation of Eq.~(\ref{eqn:pt-pt-driving}). In this case, when $v/v_T < 0.5$, we see that the $\PT$ symmetry is always broken, as is expected by the presence of edge states with pure imaginary eigenvalues. 

Similarly in Fig.~\ref{fig:pt-phases}(d), we modulate the Hamiltonian between opposing $\PT$-symmetric systems as in Eq.~(\ref{eqn:pt-pt-driving}). Note that since Fig.~\ref{fig:pt-phases}(c) and (d) have the same driving type, there is correspondence between the two graphs, which we have demonstrated by the vertical dashed line in (d) and the horizontal dashed line in (c). We observe that in this case, when $v/v_T < 0.5$, regions of unbroken $\PT$ symmetry may arise (corresponding to the black areas of the plot). Above the high driving-frequency threshold, we see that the spectrum is completely real for all coupling choices, and thus resembles the Hermitian SSH model, in accordance with the previous study \cite{Yuce2015}. We reveal that, even below this threshold, there are other regions of $\PT$-symmetry with $v/v_T < 0.5$, which arise just above the corresponding driving-frequency resonance points and which still retain localized edge states. These resonance points are highlighted by the dashed white lines.

\subsection{Analytical Result} \label{sec:analytical-floquet}

To better understand the topology and stability, we wish to analytically explore the effective Floquet Hamiltonian. Because the time dependence is a simple two-step function, the time evolution of the system over one period of driving is the product of the propagators for the static Hamiltonians associated with each step:
\begin{equation}\label{eqn:monodromy}
    G(T) = e^{-i\tilde{H}_\PT\tau}e^{-iH_\PT\tau} \,,
\end{equation}
where we introduced the shorthand $\tau \equiv T/2$.

Thus, we move to the momentum space, where each 2$\times$2 block, as in Eq.~(\ref{eqn:hptk}), of the Hamiltonian associated with momentum $k$ is independent. Therefore we may define a rotation which rotates each corresponding subspace around the $z$-axis such that we can identify the $k$-dependent rotated Hamiltonian,
\begin{equation}
    H_\pm(k) = r(k)\sigma_x \pm i\gamma\sigma_z = \vec{\lambda}_\pm(k) \cdot \vec{\sigma} \,,
\end{equation}
where $H_\PT(k) = H_+(k)$ and $\tilde{H}_\PT(k) = H_-(k)$ correspond to the two steps of the Floquet sequence, while $\vec{\lambda}_\pm$ are the vectors $[r(k), 0, \pm i\gamma]^T$ respectively (with $^T$ the vectorial transposition operation) and $\vec{\sigma} = [\sigma_x, \sigma_y, \sigma_z]^T$ is the Pauli vector such that $H_\pm = \vec{\lambda}_\pm(k)\cdot\vec{\sigma}$. We may write $\vec{\lambda}(k)_\pm = E(k)\hat{\lambda}_\pm(k)$, where $E(k) = \sqrt{r(k)^2 - \gamma^2}$ is a (possibly complex) energy eigenvalue of both $H_\pm(k)$, whereas $\hat{\lambda}_\pm(k)$ is a (possibly complex-valued) unit vector.

Dropping the argument $k$ from $E(k)$ and $\vec{\lambda}_\pm(k)$ for convenience, we may then obtain the time evolution operator for evolution of a rotated, $k$-space block, up to one driving period $T = 2\tau \equiv 2\pi/\omega$ (with driving frequency $\omega$) Eq.~(\ref{eqn:monodromy}):
\begin{align}
    \nonumber G(T) &= (\cos{E\tau} - i \sin{E\tau} \hat{\lambda}_+\cdot\vec{\sigma})(\cos{E\tau} - i \sin{E\tau} \hat{\lambda}_-\cdot\vec{\sigma}) \\
    &= \left(\cos^2{E\tau} - \sin^2{E\tau} \frac{r^2 + \gamma^2}{E^2}\right)I \nonumber \\
    &+ 2\frac{r}{E}\sin{E\tau} \left(\cos{E\tau} \hat{x} - i \frac{\gamma}{E}\sin{E\tau} \hat{y}\right)\cdot\vec{\sigma} \,,
\end{align}
where $\hat{x} = [1, 0, 0]^T$, $\hat{y} = [0, 1, 0]^T$ are unit vectors and we used the formula $(\vec{a}\cdot\vec{\sigma})(\vec{b}\cdot\vec{\sigma}) = (\vec{a}\cdot\vec{b})I + i(\vec{a}\times\vec{b})\cdot\vec{\sigma}$. From this form, we can easily identify the Floquet Hamiltonian $H_F$ and its eigenvalues (up to arbitrary terms $n\omega$ for integer $n$) $\pm\mathcal{E}$ in the form $\mathcal{E}\hat{H}_F\cdot\vec{\sigma}$, where $\mathcal{E}$ is found by solving:
\begin{equation}\label{eqn:floqeig}
    \cos(2\mathcal{E}\tau) = 1 - 2\sin^2(\mathcal{E}\tau) = 1 - 2\frac{r^2}{E^2}\sin^2{E\tau} \,,
\end{equation}
and $\hat{H}_F$ is found from
\begin{equation}\label{eqn:floqvec}
    \sin(2\mathcal{E}\tau)\hat{H}_F = 2\frac{r}{E}\sin{E\tau} \left(\cos{E\tau} \hat{x} - i \frac{\gamma}{E}\sin{E\tau} \hat{y}\right) \,,
\end{equation}
by utilizing the value of $\mathcal{E}$ found in Eq.~(\ref{eqn:floqeig}). Then the effective Floquet Hamiltonian is fully described by $H_F = \mathcal{E}\hat{H}_F$. 

By the form of Eq.~(\ref{eqn:floqvec}), we see that $H_F$ can be written
\begin{equation} \label{eqn:hf}
    H_F = c\left(\cos{E\tau}\sigma_x - i\frac{\gamma}{E}\sin{E\tau}\sigma_y\right)\,,
\end{equation}
with $c \equiv (2\mathcal{E}/\sin{2\mathcal{E}\tau})(\sin{E\tau}/E)$. Restricting the real part of $\mathcal{E}$ to $[-\omega/2, \omega/2)$, we see that there are only two cases to handle: either $\mathcal{E}$ is real, or, when $|(r/E)\sin{E\tau}| > 1$, $\mathcal{E} = \omega/2 + i\eta$ where $\eta$ is real. 

We see that when $\mathcal{E}$ is real, $c$ is real, and $H_F$ has the symmetries
\begin{align}
    \sigma_z H_F \sigma_z &= -H_F \,, \label{eqn:sublattice}\\
    \sigma_x H_F \sigma_x &= H_F^\dagger\,, \label{eqn:pseudohermitian}\\
    \sigma_y H_F \sigma_y &= -H_F^\dagger\, \label{eqn:chiral},
\end{align}
which correspond to sublattice symmetry for Eq.~(\ref{eqn:sublattice}), pseudo-Hermitian symmetry for Eq.~(\ref{eqn:pseudohermitian}), and chiral symmetry respectively for Eq.~(\ref{eqn:chiral}); and consequently, the Hamiltonian belongs to the non-Hermitian symmetry class AIII according to Ref.~\cite{Kowabata2019}.

When $\mathcal{E}$ is complex, so is $c$, and the latter two symmetries of Eqs.~(\ref{eqn:pseudohermitian}) and (\ref{eqn:chiral}) are removed. However, as previously mentioned, the solutions to Eq.~(\ref{eqn:floqeig}) also admit the solutions $\mathcal{E} + n\omega$ for any integer $n$. Rather than considering the eigenvalue pair $\pm\mathcal{E}$ which correspond to $\omega/2 + i\eta$ and $-\omega/2 - i\eta$, we may consider a solution which, for example, gives the complex conjugate pair $\omega/2 + i\eta$ and $\omega/2 - i\eta$. An alternate form of $H_F$ which takes into account this situation is
\begin{equation}
    H_{F,s} \equiv V\left[\mathcal{E}_s\sigma_z + \frac{\omega}{2} I\right]V^{-1} \,,
\end{equation}
where $\mathcal{E}_s = \mathcal{E} - \omega/2$ is the shifted eigenvalue and $V$ is the matrix of eigenvalues such that $H_F = \mathcal{E}V\sigma_zV^{-1}$. Importantly, when $\mathcal{E}$ is complex, $\mathcal{E}_s$ is pure imaginary. While this Hamiltonian is not traceless, as in Eq.~(\ref{eqn:hf}), it admits the pseudo-Hermitian symmetry of Eq.~(\ref{eqn:pseudohermitian}) even when the eigenvalues are complex. Thus, for all parameterizations of the model, the effective Hamiltonian can be seen to have this symmetry. Starting with a Floquet-modulated $\PT$-symmetric Hamiltonian, we have obtained (in the Floquet picture) a non-Hermitian Hamiltonian with pseudo-Hermitian symmetry \cite{Mostafazadeh2002}.

\subsection{Analysis of the Edge States} \label{sec:edge-states}

We also may analyze the fate of the edge states in the Floquet-driven $\PT$-SSH lattice.

\begin{figure*}
    \centering
    \includegraphics[width=\textscale]{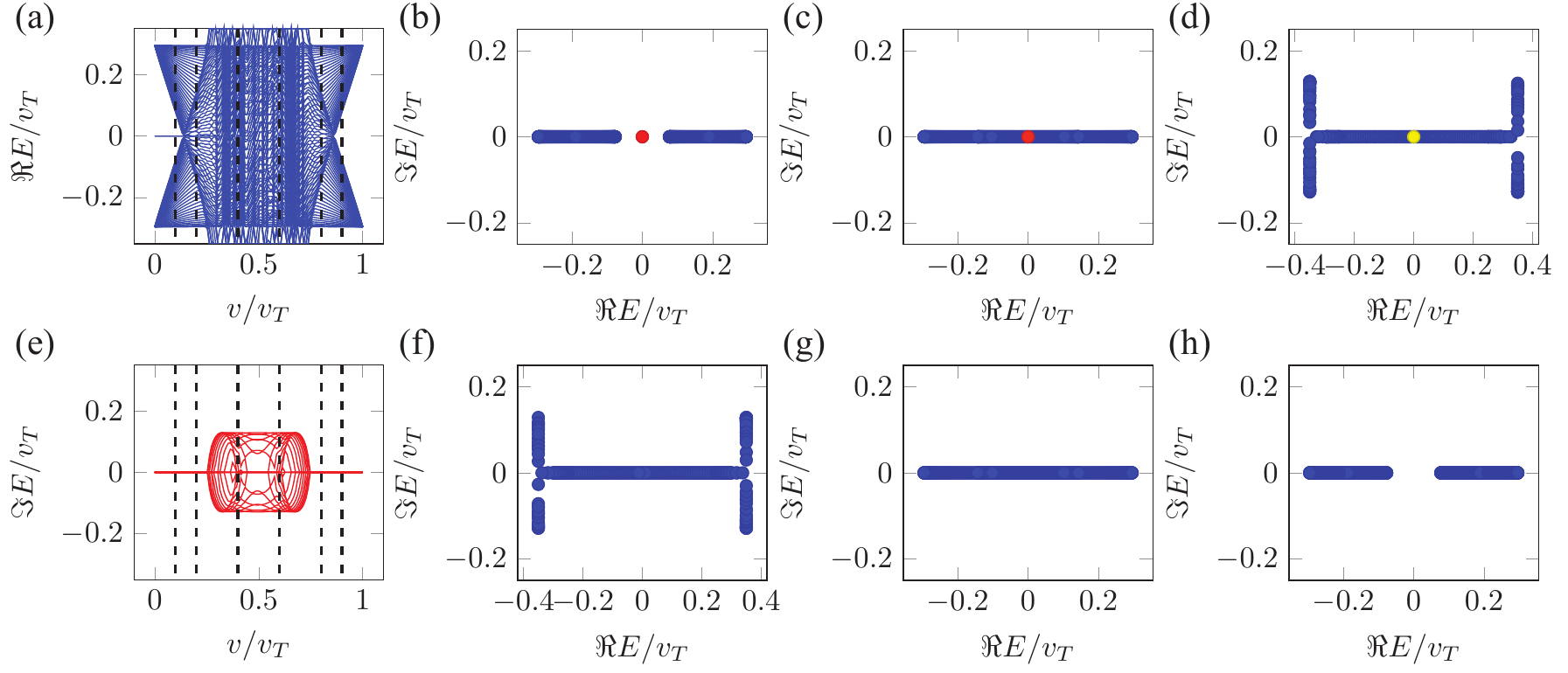}
    \caption{The complex energy spectrum of the Floquet $\PT$-symmetric system of Sec.~\ref{sec:floquet-pt} as a function of $v/v_T$, with driving frequency $\omega/v_T = 0.7$ and gain/loss rate $\gamma/v_T = 0.2$. The complex energies are plotted in the complex plane, with each marker colored by its IPR (blue at zero to red at one). In panels (b)-(d), the spectrum of the Hamiltonian is shown on the complex plane for specific configurations below the topological transition $v/v_T < 0.5$, (b) $v/v_T = 0.1$, (c) $v/v_T = 0.2$ and (d) $v/v_T = 0.4$. (f)-(h) the complex energy spectrum above the topological transition $v/v_T > 0.5$, (f) $v/v_T = 0.6$, (g) $v/v_T = 0.8$, and (h) $v/v_T = 0.9$. (a) and (e) the real and imaginary parts, respectively, of the energy bands as a function the configuration parameter $v/v_T$. The vertical dashed lines correspond to the six values of $v/v_T$ highlighted by (b)-(d) and (f)-(h).}
    \label{fig:floq-pt-ssh-bands}
\end{figure*}

For modulation between a $\PT$-symmetric Hamiltonian and a Hermitian one, as in Fig.~\ref{fig:pt-phases}(b), we find that the energy eigenvalues corresponding to the edge states lie in the complex plane, and the system never has localized edge states in the $\PT$-symmetric phase. This is a result of the inability for the imaginary parts of the associated energies to balance one another during a single oscillation. Just as in the static $\PT$-SSH case, the edge states which correspond to the resulting imaginary eigenvalues occur when $v/v_T < 0.5$ and are highly localized. In fact, the result of the static $\PT$-SSH model with gain/loss parameter $\gamma$ is recovered identically in the high-frequency regime when the Floquet driving is between $\gamma_1=2\gamma$ and $\gamma_2=0$, i.e. the average $(\gamma_1 + \gamma_2)/2$ is equal to $\gamma$.

In Fig.~\ref{fig:pt-phases}(d), we can see that in the case of $\PT$-symmetric to $\PT$-symmetric driving, the same $\PT$ phase diagram produces regions of dramatic stability within the topologically nontrivial phase. In this case, the alternation of the source and drain allows for cancellation in a single driving oscillation, i.e. the average of the gain/loss parameter over one period is zero, thus protecting the system from breaking the $\PT$ symmetry in the presence of the edge modes. In this case, in Fig.~\ref{fig:floq-pt-ssh-bands}, we see that the result of the static SSH model can be recovered, where the edge state corresponds to a real, zero-energy eigenvalue, and the system is simultaneously $\PT$-symmetric and has $v/v_T < 0.5$. 

In Fig.~\ref{fig:floq-pt-ssh-bands}(b)-(d), we observe the presence of real, mid-gap states (red/yellow). Between (b) and (c), the gap closes, but remains real, and the edge states remain localized inside the bulk spectrum. Between (d) and (f), the bulk states enter the imaginary portion of the complex plane, but again the edge states remain real. In Fig.~\ref{fig:floq-pt-ssh-bands}(f)-(h) we observe the reversal of the changes from (b)-(d), except that the edge state is removed; from (f) to (g), the bulk spectrum becomes real again, and from (g) to (h), a gap opens between the real bulk bands.

Specifically, for the complex energies depicted in Fig.~\ref{fig:floq-pt-ssh-bands}(b)-(d), the localization (IPR) of the corresponding states is indicated by the color of each marker, with the zero-energy edge states depicted with IPR $\approx 0.98$ (red) in (b), $\approx 0.88$ (red) in (c), and $\approx 0.38$ (yellow) in (d). In Fig.~\ref{fig:floq-pt-ssh-bands}(f)-(h), the highest corresponding IPR is below $0.01$.

It is important to remember that the energy spectrum shown here is restricted to values which are between $\pm\omega/2$, and that this spectrum pattern can be repeated infinitely along the real axis in both directions. We also observe that the localized, zero-energy mode is present even after the bulk gap has closed in the complex plane and exists within the bulk. 

\begin{figure}
    \centering
    \includegraphics[width=\columnwidth]{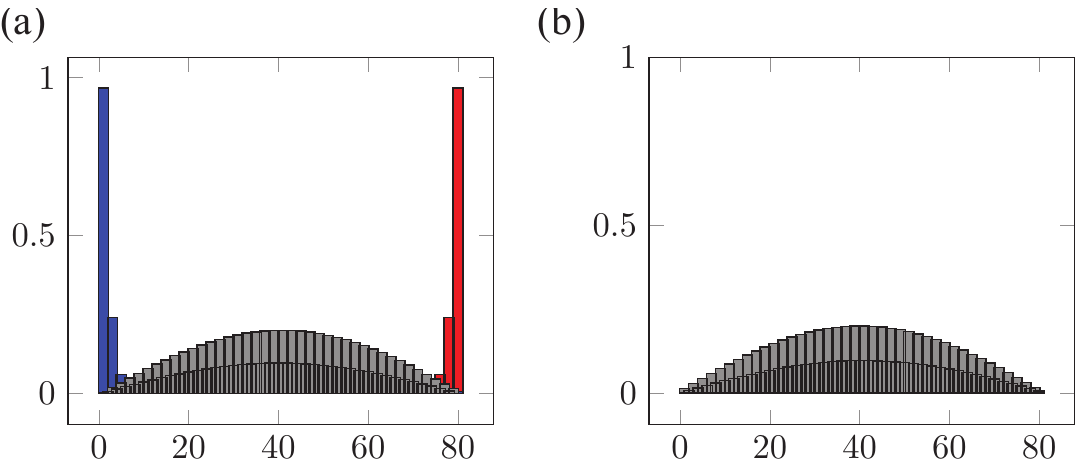}
    \caption{State amplitudes (by magnitude) of the Floquet $\PT$ SSH model. In (a), we show an example with $v/v_T = 0.2$, for which it is clear that there exist localized edge states (shown in blue and red) along with the typical bulk states (dark gray). In (b), we show an example with $v/v_T = 0.8$; we have plotted the state with the highest IPR, to demonstrate that it is not localized to either edge. In both cases, the driving frequency is $\omega/v_T = 0.7$ and the gain/loss rate is $\gamma/v_T = 0.2$}
    \label{fig:floq-pt-ssh-states} 
\end{figure}

In Fig.~\ref{fig:floq-pt-ssh-states}, we provide a specific example to demonstrate the localization of the edge states in this model. In (a), we show the site amplitudes (by magnitude) of the edge states as well as a bulk state for a realization with $v/v_T < 0.5$. Here, the localization of the left (blue) and right (red) edge states is clear, while the bulk (dark gray) remains highly delocalized. Similarly in (b), for a realization with $v/v_T > 0.5$, we show the site amplitudes for the state with the highest IPR; we specifically note the clear absence of localized edge states in this case.

\section{Conclusion} \label{sec:conclusion}

In conclusion, we have confirmed that Floquet modulation is able to provide a stabilizing effect to the static $\PT$-SSH model by stabilizing the the edge modes. The choice of modulation has nontrivial effects on the presence of the $\PT$-symmetric phase. When the modulation is kept $\PT$ symmetric for half a period and then Hermitian for the other half, in the Floquet picture, we see that, at high enough frequencies, the effective Hamiltonian has an eigenspectrum reminiscent of the static $\PT$-SSH lattice with complex edge modes. However, when we drive between a pair of time-reversed $\PT$-symmetric lattices, the high-frequency spectrum of the effective Hamiltonian resembles that of the static Hermitian SSH lattice with real and stable edge modes.

Above this high-driving frequency, the Floquet effective Hamiltonian has all real eigenvalues for all choices of lattice parameter up to a critical gain/loss rate (the $\PT$-symmetry breaking threshold). Importantly, below this high frequency regime, we find many other regions of stability. Specifically, below the first resonance frequency ($\omega/v_T = 2$) but above the second ($\omega/v_T = 2/3$), there can exist a broad portion of the $\PT$-phase diagram which is still in the $\PT$-unbroken phase and which has real, zero-energy, localized edge states.

Finally, we stress that this model is amenable to experimental implementation. Because of the simple, temporally-periodic switching used as the driving pattern, the burden of matching the gain/loss rates to a smooth, continuous function is lifted. Possibilities include purely loss-only systems where the location of the loss is alternated; after a simple imaginary gauge transformation, such systems map directly onto the one we have discussed here.

\begin{acknowledgments}
    The authors would like to acknowledge the financial support of Japan Society for the Promotion of Science (JSPS) Grants-in-Aid for Scientific Research (KAKENHI) JP19F19321 (A. H.) and JP19H00658 (N. H.). A. H. also acknowledges support as an International Research Fellow of JSPS (Postdoctoral Fellowships for Research in Japan (Standard)). Both authors would also like to thank Kohei Kowabata and Yogesh N. Joglekar for their helpful and insightful discussions and comments.
\end{acknowledgments}

\bibliography{references}

\begin{thebibliography}{33}%
\makeatletter
\providecommand \@ifxundefined [1]{%
 \@ifx{#1\undefined}
}%
\providecommand \@ifnum [1]{%
 \ifnum #1\expandafter \@firstoftwo
 \else \expandafter \@secondoftwo
 \fi
}%
\providecommand \@ifx [1]{%
 \ifx #1\expandafter \@firstoftwo
 \else \expandafter \@secondoftwo
 \fi
}%
\providecommand \natexlab [1]{#1}%
\providecommand \enquote  [1]{``#1''}%
\providecommand \bibnamefont  [1]{#1}%
\providecommand \bibfnamefont [1]{#1}%
\providecommand \citenamefont [1]{#1}%
\providecommand \href@noop [0]{\@secondoftwo}%
\providecommand \href [0]{\begingroup \@sanitize@url \@href}%
\providecommand \@href[1]{\@@startlink{#1}\@@href}%
\providecommand \@@href[1]{\endgroup#1\@@endlink}%
\providecommand \@sanitize@url [0]{\catcode `\\12\catcode `\$12\catcode
  `\&12\catcode `\#12\catcode `\^12\catcode `\_12\catcode `\%12\relax}%
\providecommand \@@startlink[1]{}%
\providecommand \@@endlink[0]{}%
\providecommand \url  [0]{\begingroup\@sanitize@url \@url }%
\providecommand \@url [1]{\endgroup\@href {#1}{\urlprefix }}%
\providecommand \urlprefix  [0]{URL }%
\providecommand \Eprint [0]{\href }%
\providecommand \doibase [0]{http://dx.doi.org/}%
\providecommand \selectlanguage [0]{\@gobble}%
\providecommand \bibinfo  [0]{\@secondoftwo}%
\providecommand \bibfield  [0]{\@secondoftwo}%
\providecommand \translation [1]{[#1]}%
\providecommand \BibitemOpen [0]{}%
\providecommand \bibitemStop [0]{}%
\providecommand \bibitemNoStop [0]{.\EOS\space}%
\providecommand \EOS [0]{\spacefactor3000\relax}%
\providecommand \BibitemShut  [1]{\csname bibitem#1\endcsname}%
\let\auto@bib@innerbib\@empty
\bibitem [{\citenamefont {Bender}\ and\ \citenamefont
  {Boettcher}(1998)}]{Bender1998}%
  \BibitemOpen
  \bibfield  {author} {\bibinfo {author} {\bibfnamefont {C.~M.}\ \bibnamefont
  {Bender}}\ and\ \bibinfo {author} {\bibfnamefont {S.}~\bibnamefont
  {Boettcher}},\ }\href {\doibase 10.1103/PhysRevLett.80.5243} {\bibfield
  {journal} {\bibinfo  {journal} {Phys. Rev. Lett.}\ }\textbf {\bibinfo
  {volume} {80}},\ \bibinfo {pages} {5243} (\bibinfo {year}
  {1998})}\BibitemShut {NoStop}%
\bibitem [{\citenamefont {Bender}\ \emph {et~al.}(2002)\citenamefont {Bender},
  \citenamefont {Brody},\ and\ \citenamefont {Jones}}]{Bender2002}%
  \BibitemOpen
  \bibfield  {author} {\bibinfo {author} {\bibfnamefont {C.~M.}\ \bibnamefont
  {Bender}}, \bibinfo {author} {\bibfnamefont {D.~C.}\ \bibnamefont {Brody}}, \
  and\ \bibinfo {author} {\bibfnamefont {H.~F.}\ \bibnamefont {Jones}},\ }\href
  {\doibase 10.1103/PhysRevLett.89.270401} {\bibfield  {journal} {\bibinfo
  {journal} {Phys. Rev. Lett.}\ }\textbf {\bibinfo {volume} {89}},\ \bibinfo
  {pages} {270401} (\bibinfo {year} {2002})}\BibitemShut {NoStop}%
\bibitem [{\citenamefont {El-Ganainy}\ \emph {et~al.}(2007)\citenamefont
  {El-Ganainy}, \citenamefont {Makris}, \citenamefont {Christodoulides},\ and\
  \citenamefont {Musslimani}}]{El-Ganainy2007}%
  \BibitemOpen
  \bibfield  {author} {\bibinfo {author} {\bibfnamefont {R.}~\bibnamefont
  {El-Ganainy}}, \bibinfo {author} {\bibfnamefont {K.~G.}\ \bibnamefont
  {Makris}}, \bibinfo {author} {\bibfnamefont {D.~N.}\ \bibnamefont
  {Christodoulides}}, \ and\ \bibinfo {author} {\bibfnamefont {Z.~H.}\
  \bibnamefont {Musslimani}},\ }\href {\doibase 10.1364/OL.32.002632}
  {\bibfield  {journal} {\bibinfo  {journal} {Opt. Lett.}\ }\textbf {\bibinfo
  {volume} {32}},\ \bibinfo {pages} {2632} (\bibinfo {year}
  {2007})}\BibitemShut {NoStop}%
\bibitem [{\citenamefont {Guo}\ \emph {et~al.}(2009)\citenamefont {Guo},
  \citenamefont {Salamo}, \citenamefont {Duchesne}, \citenamefont {Morandotti},
  \citenamefont {Volatier-Ravat}, \citenamefont {Aimez}, \citenamefont
  {Siviloglou},\ and\ \citenamefont {Christodoulides}}]{Guo2009}%
  \BibitemOpen
  \bibfield  {author} {\bibinfo {author} {\bibfnamefont {A.}~\bibnamefont
  {Guo}}, \bibinfo {author} {\bibfnamefont {G.~J.}\ \bibnamefont {Salamo}},
  \bibinfo {author} {\bibfnamefont {D.}~\bibnamefont {Duchesne}}, \bibinfo
  {author} {\bibfnamefont {R.}~\bibnamefont {Morandotti}}, \bibinfo {author}
  {\bibfnamefont {M.}~\bibnamefont {Volatier-Ravat}}, \bibinfo {author}
  {\bibfnamefont {V.}~\bibnamefont {Aimez}}, \bibinfo {author} {\bibfnamefont
  {G.~A.}\ \bibnamefont {Siviloglou}}, \ and\ \bibinfo {author} {\bibfnamefont
  {D.~N.}\ \bibnamefont {Christodoulides}},\ }\href {\doibase
  10.1103/PhysRevLett.103.093902} {\bibfield  {journal} {\bibinfo  {journal}
  {Phys. Rev. Lett.}\ }\textbf {\bibinfo {volume} {103}},\ \bibinfo {pages}
  {093902} (\bibinfo {year} {2009})}\BibitemShut {NoStop}%
\bibitem [{\citenamefont {R{\"u}ter}\ \emph {et~al.}(2010)\citenamefont
  {R{\"u}ter}, \citenamefont {Makris}, \citenamefont {El-Ganainy},
  \citenamefont {Christodoulides}, \citenamefont {Segev},\ and\ \citenamefont
  {Kip}}]{Ruter2010}%
  \BibitemOpen
  \bibfield  {author} {\bibinfo {author} {\bibfnamefont {C.~E.}\ \bibnamefont
  {R{\"u}ter}}, \bibinfo {author} {\bibfnamefont {K.~G.}\ \bibnamefont
  {Makris}}, \bibinfo {author} {\bibfnamefont {R.}~\bibnamefont {El-Ganainy}},
  \bibinfo {author} {\bibfnamefont {D.~N.}\ \bibnamefont {Christodoulides}},
  \bibinfo {author} {\bibfnamefont {M.}~\bibnamefont {Segev}}, \ and\ \bibinfo
  {author} {\bibfnamefont {D.}~\bibnamefont {Kip}},\ }\href {\doibase
  10.1038/nphys1515} {\bibfield  {journal} {\bibinfo  {journal} {Nat. Phys.}\
  }\textbf {\bibinfo {volume} {6}},\ \bibinfo {pages} {192} (\bibinfo {year}
  {2010})}\BibitemShut {NoStop}%
\bibitem [{\citenamefont {Feng}\ \emph {et~al.}(2017)\citenamefont {Feng},
  \citenamefont {El-Ganainy},\ and\ \citenamefont {Ge}}]{Feng2017}%
  \BibitemOpen
  \bibfield  {author} {\bibinfo {author} {\bibfnamefont {L.}~\bibnamefont
  {Feng}}, \bibinfo {author} {\bibfnamefont {R.}~\bibnamefont {El-Ganainy}}, \
  and\ \bibinfo {author} {\bibfnamefont {L.}~\bibnamefont {Ge}},\ }\href
  {\doibase 10.1038/s41566-017-0031-1} {\bibfield  {journal} {\bibinfo
  {journal} {Nat. Photon.}\ }\textbf {\bibinfo {volume} {11}},\ \bibinfo
  {pages} {752} (\bibinfo {year} {2017})}\BibitemShut {NoStop}%
\bibitem [{\citenamefont {El-Ganainy}\ \emph {et~al.}(2018)\citenamefont
  {El-Ganainy}, \citenamefont {Makris}, \citenamefont {Khajavikhan},
  \citenamefont {Musslimani}, \citenamefont {Rotter},\ and\ \citenamefont
  {Christodoulides}}]{El-Ganainy2018}%
  \BibitemOpen
  \bibfield  {author} {\bibinfo {author} {\bibfnamefont {R.}~\bibnamefont
  {El-Ganainy}}, \bibinfo {author} {\bibfnamefont {K.~G.}\ \bibnamefont
  {Makris}}, \bibinfo {author} {\bibfnamefont {M.}~\bibnamefont {Khajavikhan}},
  \bibinfo {author} {\bibfnamefont {Z.~H.}\ \bibnamefont {Musslimani}},
  \bibinfo {author} {\bibfnamefont {S.}~\bibnamefont {Rotter}}, \ and\ \bibinfo
  {author} {\bibfnamefont {D.~N.}\ \bibnamefont {Christodoulides}},\ }\href
  {\doibase 10.1038/nphys4323} {\bibfield  {journal} {\bibinfo  {journal} {Nat.
  Phys.}\ }\textbf {\bibinfo {volume} {14}},\ \bibinfo {pages} {11} (\bibinfo
  {year} {2018})}\BibitemShut {NoStop}%
\bibitem [{\citenamefont {\c{S}. K.~{\"O}zdemir}\ \emph
  {et~al.}(2019)\citenamefont {\c{S}. K.~{\"O}zdemir}, \citenamefont {Rotter},
  \citenamefont {Nori},\ and\ \citenamefont {Yang}}]{Ozdemir2019}%
  \BibitemOpen
  \bibfield  {author} {\bibinfo {author} {\bibnamefont {\c{S}.
  K.~{\"O}zdemir}}, \bibinfo {author} {\bibfnamefont {S.}~\bibnamefont
  {Rotter}}, \bibinfo {author} {\bibfnamefont {F.}~\bibnamefont {Nori}}, \ and\
  \bibinfo {author} {\bibfnamefont {L.}~\bibnamefont {Yang}},\ }\href {\doibase
  10.1038/s41563-019-0304-9} {\bibfield  {journal} {\bibinfo  {journal} {Nat.
  Mater.}\ }\textbf {\bibinfo {volume} {18}},\ \bibinfo {pages} {783} (\bibinfo
  {year} {2019})}\BibitemShut {NoStop}%
\bibitem [{\citenamefont {Zeuner}\ \emph {et~al.}(2015)\citenamefont {Zeuner},
  \citenamefont {Rechtsman}, \citenamefont {Plotnik}, \citenamefont {Lumer},
  \citenamefont {Nolte}, \citenamefont {Rudner}, \citenamefont {Segev},\ and\
  \citenamefont {Szameit}}]{Zeuner2015}%
  \BibitemOpen
  \bibfield  {author} {\bibinfo {author} {\bibfnamefont {J.~M.}\ \bibnamefont
  {Zeuner}}, \bibinfo {author} {\bibfnamefont {M.~C.}\ \bibnamefont
  {Rechtsman}}, \bibinfo {author} {\bibfnamefont {Y.}~\bibnamefont {Plotnik}},
  \bibinfo {author} {\bibfnamefont {Y.}~\bibnamefont {Lumer}}, \bibinfo
  {author} {\bibfnamefont {S.}~\bibnamefont {Nolte}}, \bibinfo {author}
  {\bibfnamefont {M.~S.}\ \bibnamefont {Rudner}}, \bibinfo {author}
  {\bibfnamefont {M.}~\bibnamefont {Segev}}, \ and\ \bibinfo {author}
  {\bibfnamefont {A.}~\bibnamefont {Szameit}},\ }\href {\doibase
  10.1103/PhysRevLett.115.040402} {\bibfield  {journal} {\bibinfo  {journal}
  {Phys. Rev. Lett.}\ }\textbf {\bibinfo {volume} {115}},\ \bibinfo {pages}
  {040402} (\bibinfo {year} {2015})}\BibitemShut {NoStop}%
\bibitem [{\citenamefont {Kremer}\ \emph {et~al.}(2019)\citenamefont {Kremer},
  \citenamefont {Biesenthal}, \citenamefont {Maczewsky}, \citenamefont
  {Heinrich}, \citenamefont {Thomale},\ and\ \citenamefont
  {Szameit}}]{Kremer2019}%
  \BibitemOpen
  \bibfield  {author} {\bibinfo {author} {\bibfnamefont {M.}~\bibnamefont
  {Kremer}}, \bibinfo {author} {\bibfnamefont {T.}~\bibnamefont {Biesenthal}},
  \bibinfo {author} {\bibfnamefont {L.~J.}\ \bibnamefont {Maczewsky}}, \bibinfo
  {author} {\bibfnamefont {M.}~\bibnamefont {Heinrich}}, \bibinfo {author}
  {\bibfnamefont {R.}~\bibnamefont {Thomale}}, \ and\ \bibinfo {author}
  {\bibfnamefont {A.}~\bibnamefont {Szameit}},\ }\href {\doibase
  10.1038/s41467-018-08104-x} {\bibfield  {journal} {\bibinfo  {journal} {Nat.
  Commun.}\ }\textbf {\bibinfo {volume} {10}},\ \bibinfo {pages} {435}
  (\bibinfo {year} {2019})}\BibitemShut {NoStop}%
\bibitem [{\citenamefont {Hodaei}\ \emph {et~al.}(2014)\citenamefont {Hodaei},
  \citenamefont {Miri}, \citenamefont {Heinrich}, \citenamefont
  {Christodoulides},\ and\ \citenamefont {Khajavikhan}}]{Hodaei2014}%
  \BibitemOpen
  \bibfield  {author} {\bibinfo {author} {\bibfnamefont {H.}~\bibnamefont
  {Hodaei}}, \bibinfo {author} {\bibfnamefont {M.-A.}\ \bibnamefont {Miri}},
  \bibinfo {author} {\bibfnamefont {M.}~\bibnamefont {Heinrich}}, \bibinfo
  {author} {\bibfnamefont {D.~N.}\ \bibnamefont {Christodoulides}}, \ and\
  \bibinfo {author} {\bibfnamefont {M.}~\bibnamefont {Khajavikhan}},\ }\href
  {\doibase 10.1126/science.1258480} {\bibfield  {journal} {\bibinfo  {journal}
  {Science}\ }\textbf {\bibinfo {volume} {346}},\ \bibinfo {pages} {975}
  (\bibinfo {year} {2014})}\BibitemShut {NoStop}%
\bibitem [{\citenamefont {Hodaei}\ \emph {et~al.}(2016)\citenamefont {Hodaei},
  \citenamefont {Miri}, \citenamefont {Hassan}, \citenamefont {Hayenga},
  \citenamefont {Heinrich}, \citenamefont {Christodoulides},\ and\
  \citenamefont {Khajavikhan}}]{Hodaei2016}%
  \BibitemOpen
  \bibfield  {author} {\bibinfo {author} {\bibfnamefont {H.}~\bibnamefont
  {Hodaei}}, \bibinfo {author} {\bibfnamefont {M.-A.}\ \bibnamefont {Miri}},
  \bibinfo {author} {\bibfnamefont {A.~U.}\ \bibnamefont {Hassan}}, \bibinfo
  {author} {\bibfnamefont {W.~E.}\ \bibnamefont {Hayenga}}, \bibinfo {author}
  {\bibfnamefont {M.}~\bibnamefont {Heinrich}}, \bibinfo {author}
  {\bibfnamefont {D.~N.}\ \bibnamefont {Christodoulides}}, \ and\ \bibinfo
  {author} {\bibfnamefont {M.}~\bibnamefont {Khajavikhan}},\ }\href {\doibase
  10.1002/lpor.201500292} {\bibfield  {journal} {\bibinfo  {journal} {Laser
  Photon. Rev.}\ }\textbf {\bibinfo {volume} {10}},\ \bibinfo {pages} {494}
  (\bibinfo {year} {2016})}\BibitemShut {NoStop}%
\bibitem [{\citenamefont {Schindler}\ \emph {et~al.}(2011)\citenamefont
  {Schindler}, \citenamefont {Li}, \citenamefont {Zheng}, \citenamefont
  {Ellis},\ and\ \citenamefont {Kottos}}]{Schindler2011}%
  \BibitemOpen
  \bibfield  {author} {\bibinfo {author} {\bibfnamefont {J.}~\bibnamefont
  {Schindler}}, \bibinfo {author} {\bibfnamefont {A.}~\bibnamefont {Li}},
  \bibinfo {author} {\bibfnamefont {M.~C.}\ \bibnamefont {Zheng}}, \bibinfo
  {author} {\bibfnamefont {F.~M.}\ \bibnamefont {Ellis}}, \ and\ \bibinfo
  {author} {\bibfnamefont {T.}~\bibnamefont {Kottos}},\ }\href {\doibase
  10.1103/PhysRevA.84.040101} {\bibfield  {journal} {\bibinfo  {journal} {Phys.
  Rev. A}\ }\textbf {\bibinfo {volume} {84}},\ \bibinfo {pages} {040101(R)}
  (\bibinfo {year} {2011})}\BibitemShut {NoStop}%
\bibitem [{\citenamefont {Bender}\ \emph {et~al.}(2013)\citenamefont {Bender},
  \citenamefont {Berntson}, \citenamefont {Parker},\ and\ \citenamefont
  {Samuel}}]{Bender2013}%
  \BibitemOpen
  \bibfield  {author} {\bibinfo {author} {\bibfnamefont {C.~M.}\ \bibnamefont
  {Bender}}, \bibinfo {author} {\bibfnamefont {B.~K.}\ \bibnamefont
  {Berntson}}, \bibinfo {author} {\bibfnamefont {D.}~\bibnamefont {Parker}}, \
  and\ \bibinfo {author} {\bibfnamefont {E.}~\bibnamefont {Samuel}},\ }\href
  {\doibase 10.1119/1.4789549} {\bibfield  {journal} {\bibinfo  {journal} {Am.
  J. Phys.}\ }\textbf {\bibinfo {volume} {81}},\ \bibinfo {pages} {173}
  (\bibinfo {year} {2013})}\BibitemShut {NoStop}%
\bibitem [{\citenamefont {Zhang}\ \emph {et~al.}(2016)\citenamefont {Zhang},
  \citenamefont {Zhang}, \citenamefont {Sheng}, \citenamefont {Yang},
  \citenamefont {Miri}, \citenamefont {Christodoulides}, \citenamefont {He},
  \citenamefont {Zhang},\ and\ \citenamefont {Xiao}}]{Zhang2016}%
  \BibitemOpen
  \bibfield  {author} {\bibinfo {author} {\bibfnamefont {Z.}~\bibnamefont
  {Zhang}}, \bibinfo {author} {\bibfnamefont {Y.}~\bibnamefont {Zhang}},
  \bibinfo {author} {\bibfnamefont {J.}~\bibnamefont {Sheng}}, \bibinfo
  {author} {\bibfnamefont {L.}~\bibnamefont {Yang}}, \bibinfo {author}
  {\bibfnamefont {M.-A.}\ \bibnamefont {Miri}}, \bibinfo {author}
  {\bibfnamefont {D.~N.}\ \bibnamefont {Christodoulides}}, \bibinfo {author}
  {\bibfnamefont {B.}~\bibnamefont {He}}, \bibinfo {author} {\bibfnamefont
  {Y.}~\bibnamefont {Zhang}}, \ and\ \bibinfo {author} {\bibfnamefont
  {M.}~\bibnamefont {Xiao}},\ }\href {\doibase 10.1103/PhysRevLett.117.123601}
  {\bibfield  {journal} {\bibinfo  {journal} {Phys. Rev. Lett.}\ }\textbf
  {\bibinfo {volume} {117}},\ \bibinfo {pages} {123601} (\bibinfo {year}
  {2016})}\BibitemShut {NoStop}%
\bibitem [{\citenamefont {Rudner}\ and\ \citenamefont
  {Levitov}(2009)}]{Rudner2009}%
  \BibitemOpen
  \bibfield  {author} {\bibinfo {author} {\bibfnamefont {M.~S.}\ \bibnamefont
  {Rudner}}\ and\ \bibinfo {author} {\bibfnamefont {L.~S.}\ \bibnamefont
  {Levitov}},\ }\href {\doibase 10.1103/PhysRevLett.102.065703} {\bibfield
  {journal} {\bibinfo  {journal} {Phys. Rev. Lett.}\ }\textbf {\bibinfo
  {volume} {102}},\ \bibinfo {pages} {065703} (\bibinfo {year}
  {2009})}\BibitemShut {NoStop}%
\bibitem [{\citenamefont {Su}\ \emph {et~al.}(1979)\citenamefont {Su},
  \citenamefont {Schrieffer},\ and\ \citenamefont {Heeger}}]{Su1979}%
  \BibitemOpen
  \bibfield  {author} {\bibinfo {author} {\bibfnamefont {W.~P.}\ \bibnamefont
  {Su}}, \bibinfo {author} {\bibfnamefont {J.~R.}\ \bibnamefont {Schrieffer}},
  \ and\ \bibinfo {author} {\bibfnamefont {A.~J.}\ \bibnamefont {Heeger}},\
  }\href {\doibase 10.1103/PhysRevLett.42.1698} {\bibfield  {journal} {\bibinfo
   {journal} {Phys. Rev. Lett.}\ }\textbf {\bibinfo {volume} {42}},\ \bibinfo
  {pages} {1698} (\bibinfo {year} {1979})}\BibitemShut {NoStop}%
\bibitem [{\citenamefont {Asb\'oth}\ \emph {et~al.}(2016)\citenamefont
  {Asb\'oth}, \citenamefont {Oroszl\'any},\ and\ \citenamefont
  {P\'alyi}}]{Asboth2016}%
  \BibitemOpen
  \bibfield  {author} {\bibinfo {author} {\bibfnamefont {J.~K.}\ \bibnamefont
  {Asb\'oth}}, \bibinfo {author} {\bibfnamefont {L.}~\bibnamefont
  {Oroszl\'any}}, \ and\ \bibinfo {author} {\bibfnamefont {A.}~\bibnamefont
  {P\'alyi}},\ }\href {\doibase 10.1007/978-3-319-25607-8} {\emph {\bibinfo
  {title} {A Short Course on Topological Insulators}}},\ \bibinfo {series}
  {Lecture Notes in Physics}, Vol.\ \bibinfo {volume} {919}\ (\bibinfo
  {publisher} {Springer, Cham},\ \bibinfo {year} {2016})\BibitemShut {NoStop}%
\bibitem [{\citenamefont {Hu}\ and\ \citenamefont {Hughes}(2011)}]{Hu2011}%
  \BibitemOpen
  \bibfield  {author} {\bibinfo {author} {\bibfnamefont {Y.~C.}\ \bibnamefont
  {Hu}}\ and\ \bibinfo {author} {\bibfnamefont {T.~L.}\ \bibnamefont
  {Hughes}},\ }\href {\doibase 10.1103/PhysRevB.84.153101} {\bibfield
  {journal} {\bibinfo  {journal} {Phys. Rev. B}\ }\textbf {\bibinfo {volume}
  {84}},\ \bibinfo {pages} {153101} (\bibinfo {year} {2011})}\BibitemShut
  {NoStop}%
\bibitem [{\citenamefont {Liang}\ and\ \citenamefont
  {Huang}(2013)}]{Liang2013}%
  \BibitemOpen
  \bibfield  {author} {\bibinfo {author} {\bibfnamefont {S.-D.}\ \bibnamefont
  {Liang}}\ and\ \bibinfo {author} {\bibfnamefont {G.-Y.}\ \bibnamefont
  {Huang}},\ }\href {\doibase 10.1103/PhysRevA.87.012118} {\bibfield  {journal}
  {\bibinfo  {journal} {Phys. Rev. A}\ }\textbf {\bibinfo {volume} {87}},\
  \bibinfo {pages} {012118} (\bibinfo {year} {2013})}\BibitemShut {NoStop}%
\bibitem [{\citenamefont {Harter}\ \emph {et~al.}(2018)\citenamefont {Harter},
  \citenamefont {Saxena},\ and\ \citenamefont {Joglekar}}]{Harter2018}%
  \BibitemOpen
  \bibfield  {author} {\bibinfo {author} {\bibfnamefont {A.~K.}\ \bibnamefont
  {Harter}}, \bibinfo {author} {\bibfnamefont {A.}~\bibnamefont {Saxena}}, \
  and\ \bibinfo {author} {\bibfnamefont {Y.~N.}\ \bibnamefont {Joglekar}},\
  }\href {\doibase 10.1038/s41598-018-30344-6} {\bibfield  {journal} {\bibinfo
  {journal} {Sci. Rep.}\ }\textbf {\bibinfo {volume} {8}},\ \bibinfo {pages}
  {12065} (\bibinfo {year} {2018})}\BibitemShut {NoStop}%
\bibitem [{\citenamefont {Lieu}(2018)}]{Lieu2018}%
  \BibitemOpen
  \bibfield  {author} {\bibinfo {author} {\bibfnamefont {S.}~\bibnamefont
  {Lieu}},\ }\href {\doibase 10.1103/PhysRevB.97.045106} {\bibfield  {journal}
  {\bibinfo  {journal} {Phys. Rev. B}\ }\textbf {\bibinfo {volume} {97}},\
  \bibinfo {pages} {045106} (\bibinfo {year} {2018})}\BibitemShut {NoStop}%
\bibitem [{\citenamefont {Shirley}(1965)}]{Shirley1965}%
  \BibitemOpen
  \bibfield  {author} {\bibinfo {author} {\bibfnamefont {J.~H.}\ \bibnamefont
  {Shirley}},\ }\href {\doibase 10.1103/PhysRev.138.B979} {\bibfield  {journal}
  {\bibinfo  {journal} {Phys. Rev.}\ }\textbf {\bibinfo {volume} {138}},\
  \bibinfo {pages} {B979} (\bibinfo {year} {1965})}\BibitemShut {NoStop}%
\bibitem [{\citenamefont {Joglekar}\ \emph {et~al.}(2014)\citenamefont
  {Joglekar}, \citenamefont {Marathe}, \citenamefont {Durganandini},\ and\
  \citenamefont {Pathak}}]{Joglekar2014}%
  \BibitemOpen
  \bibfield  {author} {\bibinfo {author} {\bibfnamefont {Y.~N.}\ \bibnamefont
  {Joglekar}}, \bibinfo {author} {\bibfnamefont {R.}~\bibnamefont {Marathe}},
  \bibinfo {author} {\bibfnamefont {P.}~\bibnamefont {Durganandini}}, \ and\
  \bibinfo {author} {\bibfnamefont {R.~K.}\ \bibnamefont {Pathak}},\ }\href
  {\doibase 10.1103/PhysRevA.90.040101} {\bibfield  {journal} {\bibinfo
  {journal} {Phys. Rev. A}\ }\textbf {\bibinfo {volume} {90}},\ \bibinfo
  {pages} {040101(R)} (\bibinfo {year} {2014})}\BibitemShut {NoStop}%
\bibitem [{\citenamefont {Lee}\ and\ \citenamefont {Joglekar}(2015)}]{Lee2015}%
  \BibitemOpen
  \bibfield  {author} {\bibinfo {author} {\bibfnamefont {T.~E.}\ \bibnamefont
  {Lee}}\ and\ \bibinfo {author} {\bibfnamefont {Y.~N.}\ \bibnamefont
  {Joglekar}},\ }\href {\doibase 10.1103/PhysRevA.92.042103} {\bibfield
  {journal} {\bibinfo  {journal} {Phys. Rev. A}\ }\textbf {\bibinfo {volume}
  {92}},\ \bibinfo {pages} {042103} (\bibinfo {year} {2015})}\BibitemShut
  {NoStop}%
\bibitem [{\citenamefont {Li}\ \emph {et~al.}(2019)\citenamefont {Li},
  \citenamefont {Harter}, \citenamefont {Liu}, \citenamefont {de~Melo},
  \citenamefont {Joglekar},\ and\ \citenamefont {Luo}}]{Li2019}%
  \BibitemOpen
  \bibfield  {author} {\bibinfo {author} {\bibfnamefont {J.}~\bibnamefont
  {Li}}, \bibinfo {author} {\bibfnamefont {A.~K.}\ \bibnamefont {Harter}},
  \bibinfo {author} {\bibfnamefont {J.}~\bibnamefont {Liu}}, \bibinfo {author}
  {\bibfnamefont {L.}~\bibnamefont {de~Melo}}, \bibinfo {author} {\bibfnamefont
  {Y.~N.}\ \bibnamefont {Joglekar}}, \ and\ \bibinfo {author} {\bibfnamefont
  {L.}~\bibnamefont {Luo}},\ }\href {\doibase 10.1038/s41467-019-08596-1}
  {\bibfield  {journal} {\bibinfo  {journal} {Nat. Commun.}\ }\textbf {\bibinfo
  {volume} {10}},\ \bibinfo {pages} {855} (\bibinfo {year} {2019})}\BibitemShut
  {NoStop}%
\bibitem [{\citenamefont {de~J.~Le\'on-Montiel}\ \emph
  {et~al.}(2018)\citenamefont {de~J.~Le\'on-Montiel}, \citenamefont
  {Quiroz-Ju\'arez}, \citenamefont {Dom\'inguez-Ju\'arez}, \citenamefont
  {Quintero-Torres}, \citenamefont {Arag\'on}, \citenamefont {Harter},\ and\
  \citenamefont {Joglekar}}]{Leon-Montiel2018}%
  \BibitemOpen
  \bibfield  {author} {\bibinfo {author} {\bibfnamefont {R.}~\bibnamefont
  {de~J.~Le\'on-Montiel}}, \bibinfo {author} {\bibfnamefont {M.~A.}\
  \bibnamefont {Quiroz-Ju\'arez}}, \bibinfo {author} {\bibfnamefont {J.~L.}\
  \bibnamefont {Dom\'inguez-Ju\'arez}}, \bibinfo {author} {\bibfnamefont
  {R.}~\bibnamefont {Quintero-Torres}}, \bibinfo {author} {\bibfnamefont
  {J.~L.}\ \bibnamefont {Arag\'on}}, \bibinfo {author} {\bibfnamefont {A.~K.}\
  \bibnamefont {Harter}}, \ and\ \bibinfo {author} {\bibfnamefont {Y.~N.}\
  \bibnamefont {Joglekar}},\ }\href {\doibase 10.1038/s42005-018-0087-3}
  {\bibfield  {journal} {\bibinfo  {journal} {Commun. Phys.}\ }\textbf
  {\bibinfo {volume} {1}},\ \bibinfo {pages} {88} (\bibinfo {year}
  {2018})}\BibitemShut {NoStop}%
\bibitem [{\citenamefont {Yuce}(2015)}]{Yuce2015}%
  \BibitemOpen
  \bibfield  {author} {\bibinfo {author} {\bibfnamefont {C.}~\bibnamefont
  {Yuce}},\ }\href {\doibase 10.1140/epjd/e2015-60220-7} {\bibfield  {journal}
  {\bibinfo  {journal} {Eur. Phys. J. D}\ }\textbf {\bibinfo {volume} {69}},\
  \bibinfo {pages} {184} (\bibinfo {year} {2015})}\BibitemShut {NoStop}%
\bibitem [{\citenamefont {Kitagawa}\ \emph {et~al.}(2010)\citenamefont
  {Kitagawa}, \citenamefont {Berg}, \citenamefont {Rudner},\ and\ \citenamefont
  {Demler}}]{Kitagawa2010}%
  \BibitemOpen
  \bibfield  {author} {\bibinfo {author} {\bibfnamefont {T.}~\bibnamefont
  {Kitagawa}}, \bibinfo {author} {\bibfnamefont {E.}~\bibnamefont {Berg}},
  \bibinfo {author} {\bibfnamefont {M.}~\bibnamefont {Rudner}}, \ and\ \bibinfo
  {author} {\bibfnamefont {E.}~\bibnamefont {Demler}},\ }\href {\doibase
  10.1103/PhysRevB.82.235114} {\bibfield  {journal} {\bibinfo  {journal} {Phys.
  Rev. B}\ }\textbf {\bibinfo {volume} {82}},\ \bibinfo {pages} {235114}
  (\bibinfo {year} {2010})}\BibitemShut {NoStop}%
\bibitem [{\citenamefont {Rechtsman}\ \emph {et~al.}(2013)\citenamefont
  {Rechtsman}, \citenamefont {Zeuner}, \citenamefont {Plotnik}, \citenamefont
  {Lumer}, \citenamefont {Podolsky}, \citenamefont {Dreisow}, \citenamefont
  {Nolte}, \citenamefont {Segev},\ and\ \citenamefont
  {Szameit}}]{Rechtsman2013}%
  \BibitemOpen
  \bibfield  {author} {\bibinfo {author} {\bibfnamefont {M.~C.}\ \bibnamefont
  {Rechtsman}}, \bibinfo {author} {\bibfnamefont {J.~M.}\ \bibnamefont
  {Zeuner}}, \bibinfo {author} {\bibfnamefont {Y.}~\bibnamefont {Plotnik}},
  \bibinfo {author} {\bibfnamefont {Y.}~\bibnamefont {Lumer}}, \bibinfo
  {author} {\bibfnamefont {D.}~\bibnamefont {Podolsky}}, \bibinfo {author}
  {\bibfnamefont {F.}~\bibnamefont {Dreisow}}, \bibinfo {author} {\bibfnamefont
  {S.}~\bibnamefont {Nolte}}, \bibinfo {author} {\bibfnamefont
  {M.}~\bibnamefont {Segev}}, \ and\ \bibinfo {author} {\bibfnamefont
  {A.}~\bibnamefont {Szameit}},\ }\href {\doibase 10.1038/nature12066}
  {\bibfield  {journal} {\bibinfo  {journal} {Nature}\ }\textbf {\bibinfo
  {volume} {496}},\ \bibinfo {pages} {196} (\bibinfo {year}
  {2013})}\BibitemShut {NoStop}%
\bibitem [{\citenamefont {Rudner}\ \emph {et~al.}(2013)\citenamefont {Rudner},
  \citenamefont {Lindner}, \citenamefont {Berg},\ and\ \citenamefont
  {Levin}}]{Rudner2013}%
  \BibitemOpen
  \bibfield  {author} {\bibinfo {author} {\bibfnamefont {M.~S.}\ \bibnamefont
  {Rudner}}, \bibinfo {author} {\bibfnamefont {N.~H.}\ \bibnamefont {Lindner}},
  \bibinfo {author} {\bibfnamefont {E.}~\bibnamefont {Berg}}, \ and\ \bibinfo
  {author} {\bibfnamefont {M.}~\bibnamefont {Levin}},\ }\href {\doibase
  10.1103/PhysRevX.3.031005} {\bibfield  {journal} {\bibinfo  {journal} {Phys.
  Rev. X}\ }\textbf {\bibinfo {volume} {3}},\ \bibinfo {pages} {031005}
  (\bibinfo {year} {2013})}\BibitemShut {NoStop}%
\bibitem [{\citenamefont {Kawabata}\ \emph {et~al.}(2019)\citenamefont
  {Kawabata}, \citenamefont {Shiozaki}, \citenamefont {Ueda},\ and\
  \citenamefont {Sato}}]{Kowabata2019}%
  \BibitemOpen
  \bibfield  {author} {\bibinfo {author} {\bibfnamefont {K.}~\bibnamefont
  {Kawabata}}, \bibinfo {author} {\bibfnamefont {K.}~\bibnamefont {Shiozaki}},
  \bibinfo {author} {\bibfnamefont {M.}~\bibnamefont {Ueda}}, \ and\ \bibinfo
  {author} {\bibfnamefont {M.}~\bibnamefont {Sato}},\ }\href {\doibase
  10.1103/PhysRevX.9.041015} {\bibfield  {journal} {\bibinfo  {journal} {Phys.
  Rev. X}\ }\textbf {\bibinfo {volume} {9}},\ \bibinfo {pages} {041015}
  (\bibinfo {year} {2019})}\BibitemShut {NoStop}%
\bibitem [{\citenamefont {Mostafazadeh}(2002)}]{Mostafazadeh2002}%
  \BibitemOpen
  \bibfield  {author} {\bibinfo {author} {\bibfnamefont {A.}~\bibnamefont
  {Mostafazadeh}},\ }\href {\doibase 10.1063/1.1418246} {\bibfield  {journal}
  {\bibinfo  {journal} {J. Math. Phys.}\ }\textbf {\bibinfo {volume} {43}},\
  \bibinfo {pages} {205} (\bibinfo {year} {2002})}\BibitemShut {NoStop}%
\end{thebibliography}%

\end{document}